\documentclass[aps,prb,twocolumn,showpacs]{revtex4}
\usepackage[dvips]{graphicx}

\begin{document}

\title{{\it s}-Wave-Like excitation in the superconducting state of electron-doped cuprates with {\it d}-wave pairing}
\author{Qingshan Yuan,$^1$ Xin-Zhong Yan,$^{1,2}$ and C. S. Ting$^1$}
\address{$^1$Texas Center for Superconductivity
and Department of Physics, University of Houston, Houston, TX 77204\\
$^2$Institute of Physics, Chinese Academy of Sciences, P. O. Box 603,
Beijing 100080, China}

\begin{abstract}
An intrinsic physical mechanism, based on the doping evolution of the Fermi surface (FS), is explored to reconcile the contradictory experimental results on the superconducting (SC) pairing symmetry in electron-doped cuprates.
It is argued that the FS pocket around $(\pi/2,\pi/2)$ has not yet formed until doping reaches about the optimal value. Therefore, in the underdoped regime,
even if the SC order parameter is $d$ wave which vanishes along the line $k_x=k_y$,
the quasiparticle excitation gap is still finite and looks $s$-wave-like
due to the absence of the FS across that line. This makes it possible, with 
$d$-wave SC pairing, to understand those experiments which evidenced the 
$s$-wave quasiparticle excitation. An explicit theory with consideration of
both antiferromagnetic and SC orders is implemented to exhibit
the FS evolution from underdoping to overdoping and associated with it
the variations of the quasiparticle property, electronic density of states,
and low temperature dependences of the physical quantities heat capacity and superfluid density.
\end{abstract}

\maketitle
\section{Introduction}
\begin{figure}[ht]
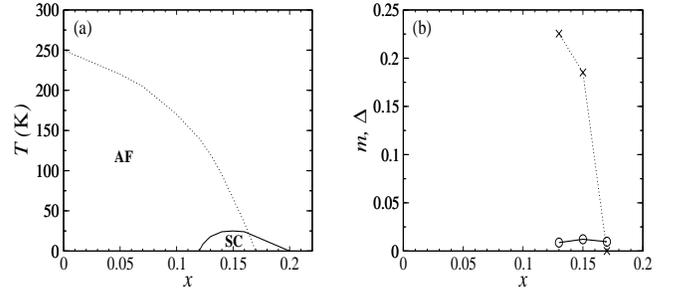

\begin{center}
\includegraphics[width=4.cm,height=3.8cm,clip]{PD.eps}\hspace*{4mm}
\includegraphics[width=4.cm,height=3.8cm,clip]{OP.eps}
\end{center}
\caption{(a) The schematic phase diagram for electron-doped cuprates Nd(Pr)$_{2-x}$Ce$_x$CuO$_4$. The AF order persists beyond optimal doping
$x=0.15$.\cite{Uefuji02,Kang,Dagan,Zimmers} (b) The calculated order parameters
$m$ (``$\times$'') and $\Delta$ (``$\circ$'') at $T=0$ for three doping levels $x=0.13,\ 0.15$ and $0.17$ from the phenomenological $t$-$t'$-$t''$-$U$-$V$ model to reproduce the experimental features. The lines are only for guiding the eyes.}
\label{Fig:PD}
\end{figure}

The electron-doped (e-doped) cuprate superconductors, typically Nd$_{2-x}$Ce$_x$CuO$_4$ (NCCO) and Pr$_{2-x}$Ce$_x$CuO$_4$ (PCCO), are an important component in the study of high-$T_c$ superconductivity.
Compared to hole-doped cuprates, the e-doped ones have a much more robust antiferromagnetic (AF) phase and a much narrower superconducting (SC) phase. Moreover, there is strong evidence for the coexistence of both AF and SC orders around optimal doping $x=0.15$,\cite{Uefuji02,Kang} although this issue is
still under debate.\cite{Motoyama}
The transport measurements on PCCO \cite{Dagan,Zimmers} suggest a quantum phase transition from antiferromagnetism to paramagnetism at $x\simeq 0.17$.
The schematic phase diagram is plotted in Fig.~\ref{Fig:PD}(a).

As the fundamental topic, the SC pairing symmetry has been intensely studied.
While the {\it s}-wave symmetry was supported by the measurements like
penetration depth \cite{Alff,Kim} and tunneling,\cite{Kashiwaya,Shan,Chen} 
the {\it d}-wave symmetry was suggested by the seemingly more 
extensive measurements including phase-sensitive Josephson junctions, \cite{Tsuei,Ariando} angle-resolved photoemission spectroscopy (ARPES), \cite{Armitage01} specific heat,\cite{Yu} and also penetration depth. \cite{Kokales,Snezhko} In addition, there is other evidence to show
the so-called nonmonotonic {\it d}-wave behavior \cite{Blumberg,Matsui95}
or a transition from $d$- to $s$-wave symmetry \cite{Biswas,Skinta}
with increasing doping.
Thus the experimental data are rather controversial to support the $s$-wave
or essentially $d$-wave (no matter monotonic or not) pairing. In particular,
a puzzle arised in the penetration depth measurements that
different results were obtained by different groups and even the same group.
For example, a finite SC gap was inferred by Kim {\it et al.} \cite{Kim}
over a wide range of doping levels from underdoping to overdoping,
even though a transition from $d$- to $s$-wave behavior with increasing doping 
was claimed earlier by the same group.\cite{Skinta}
On the other hand, completely contrary result, i.e., $d$-wave behavior
at various doping levels, was reported by Snezhko et al.\cite{Snezhko}
Why are the experimental data so contradictory? Perhaps it is partly due to differences between fabricated samples. However, in this paper we intend to
explore an intrinsic physical mechanism, based on the doping evolution of the
Fermi surface (FS), to understand the diverse experimental results. With it the apparent discrepancies might be largely reconciled.

To expound the mechanism, we first examine the FS structure of
e-doped cuprates. The ARPES measurements on NCCO \cite{Armitage02,Matsui94}
have revealed the intriguing FS evolution with doping.
It was found that at low doping $x=0.04$ only a small FS pocket appears around $(\pi,0)$. With increasing doping new spectral intensity around $(\pi/2,\pi/2)$
begins to emerge and becomes enough strong at optimal doping $x=0.15$,
indicating the formation of the new FS pocket around $(\pi/2,\pi/2)$.
The ARPES data have been explained within the two-band model
in consideration of the AF order \cite{Kusko,Yuan,Yan} or by
numerically considering the AF short-range correlation.\cite{Tremblay}
According to the two-band analysis, at low doping only the upper AF band is crossed by the Fermi level around $(\pi,0)$, leading to the small FS pocket around $(\pi,0)$. With increasing doping the AF order weakens so that the lower band, with the maximum at $(\pi/2,\pi/2)$, approaches the Fermi level and
contributes the new spectral intensity around $(\pi/2,\pi/2)$. The question is at which doping the lower band will be crossed by the Fermi level. As seen from the ARPES data on NCCO at $x=0.13$ (Fig.~2 of Ref.~\onlinecite{Matsui94}),
the lower band is still away from the Fermi level, although it is close to the
latter and thus contributes weak spectral intensity around $(\pi/2,\pi/2)$
in the intensity map (Fig.~1 of Ref.~\onlinecite{Matsui94}).
In view that the lower band lifts continuously with increasing doping,
we naturally imagine that it does not touch the Fermi level until $x\simeq 0.15$. That is to say, in nearly the whole underdoped regime $(x<0.15)$,
the real FS pocket around $(\pi/2,\pi/2)$ has not yet formed.

On the other hand, as far as we know, many experimental measurements to study the SC pairing symmetry only give direct information about the quasiparticle excitation gap, and not the SC order parameter. For example, from the low temperature dependence of the penetration depth (exponential function or power law), one can only infer whether the quasiparticle excitation has a finite gap or not.
It must be noted that a finite quasiparticle excitation gap,
which was actually obtained by e.g. Kim {\it et al.},\cite{Kim}
does not necessarily mean a nodeless SC order parameter. Here our essential idea is: The quasiparticle excitation gap may be finite and look
{\it s}-wave-like, but the SC order parameter is {\it d}-wave and has nodes. This possibility becomes
true if the nodal line of the SC order parameter does not cross the FS.
Such a case is exactly exhibited above in the underdoped regime
where the FS is lacking around $(\pi/2,\pi/2)$. Therefore, even if the SC order 
parameter is {\it d}-wave which vanishes along the line $k_x=k_y$,
the quasiparticle excitation is still gapped due to the absence of the FS across that line.

As indicated above, it is of general interest to investigate how the quasiparticle property varies with the intrinsic FS evolution with doping in e-doped cuprates.
This guides us to conceive the following scenario to largely reconcile the
experimental discrepancies. Simply we assume that the SC order parameter
is $d$-wave at all doping levels. This is consistent with the phase sensitive experiments \cite{Tsuei,Ariando} on optimally doped NCCO and PCCO and overdoped NCCO as well as the ARPES measurements \cite{Armitage01} on optimally doped NCCO. In view of the FS evolution, we expect that the quasiparticle
excitation will have a finite, i.e., {\it s}-wave-like gap in the underdoped regime and zero gap in the optimally and overdoped regime. Then those penetration depth measurements to support the {\it s}-wave gap might be consistently understood if nominally optimally or overdoped samples in those measurements are in reality underdoped.

The paper is organized as follows.
In the next section the model Hamiltonian is introduced and the diagonalization
is performed. In Sec.~III, the FS evolution from underdoping to overdoping,
and simultaneously the variation of
the quasiparticle property and electronic density of states (DOS), are exhibited. A remarkable new finding, i.e., a double-peak feature, in the DOS
at optimal doping is discovered.
In sections IV and V the physical quantities heat capacity and superfluid density are calculated, respectively, and their low temperature dependences
are discussed. Conclusive remarks are given in Sec.~VI. Two appendices A and B
are supplemented to describe the details of Hamiltonian diagonalization and derivation of the electronic spectral function, respectively. 

\section{Model Hamiltonian and diagonalization}
Our idea itself is independent of the concrete model calculations as long as
the argued FS structures at different doping levels are recognized.
The microscopic Hubbard and $t$-$J$ type models, as extensively applied
to e-doped cuprates, are difficult \cite{Senechal,Li} to reproduce the global phase diagram. Alternatively, we take as our starting point the experimental phase diagram and for convenience adopt the phenomenological $t$-$t'$-$t''$-$U$-$V$ model with repulsive on-site $U$ and attractive intersite $V$ to simulate the experimental phases. For the $V$-term $-V\sum_{\langle ij\rangle}n_i n_j$,
only the SC pairing interaction is retained.
The model Hamiltonian reads as follows after mean-field decoupling
(without irrelevant constant)
\begin{eqnarray}
H & = & -t\sum_{\langle ij\rangle \sigma}(c_{i\sigma}^{\dagger}c_{j\sigma}+{\rm H.c.})
-t'\sum_{\langle ij\rangle'\sigma}(c_{i\sigma}^{\dagger}c_{j\sigma}+{\rm H.c.})
\nonumber\\
& & -t''\sum_{\langle ij\rangle'' \sigma}(c_{i\sigma}^{\dagger}c_{j\sigma}+{\rm H.c.}) - \sum_{i\sigma} [Um(-1)^i \sigma +\mu] c_{i\sigma}^{\dagger}c_{i\sigma}
\nonumber\\
& & -(V/2)\sum_{\langle ij\rangle}[\Delta_{ij}^*(c_{i\uparrow}c_{j\downarrow}-
c_{i\downarrow}c_{j\uparrow})+{\rm H.c.}] \nonumber\\
& & + NUm^2+(V/2)\sum_{\langle ij\rangle}|\Delta_{ij}|^2,\label{H}
\end{eqnarray}
where $\langle\rangle,\ \langle\rangle',\ \langle\rangle''$
represent the nearest neighbor (n.n.), second n.n., and third n.n. sites, 
respectively, $m=(-1)^i \langle c_{i\uparrow}^{\dagger}c_{i\uparrow}-
c_{i\downarrow}^{\dagger}c_{i\downarrow}\rangle/2$ is the AF order parameter,
$\Delta_{ij}=\langle c_{i\uparrow}c_{j\downarrow}-c_{i\downarrow}c_{j\uparrow}
\rangle=2\langle c_{i\uparrow}c_{j\downarrow}\rangle$ is the pairing order parameter, $N$ is the total number of lattice sites, and the rest of the notation is standard. As mentioned above, we assume throughout the doping range that the pairing symmetry is standard $d$ wave, i.e., 
$\Delta_{ij}=\Delta\ (-\Delta)$ for bond $\langle ij\rangle$
along $x\ (y)$ direction. We set $t$ as the energy unit and 
$k_B=\hbar=a=1$ ($a$: lattice constant). Typical values $t'=-0.3$ and $t''=0.2$
are adopted.

To be transparent, the original lattice is divided into two sublattices D and E,
with net spin orientation $\uparrow$ and $\downarrow$, respectively.
Correspondingly new fermionic operators $d$ and $e$ for the two sublattices
are introduced, 
i.e., $c_{i\sigma}=d_{i\sigma}\ (e_{i\sigma})$ for $i\in {\rm D}\ ({\rm E})$.
Then the Hamiltonian (\ref{H}) can be written in momentum space as
\begin{equation}
H=\sum'_k [\hat{\Psi}_k^{\dagger} \hat{H}_k \hat{\Psi}_k 
+ 2(\varepsilon'_k-\mu)]+N(Um^2+V\Delta^2),\label{HM}
\end{equation}
where the four-component operator $\hat{\Psi}_k^{\dagger}$ is defined as
$$\hat{\Psi}_k^{\dagger}=(d_{k\uparrow}^{\dagger}\ e_{-k\downarrow}\ 
e_{k\uparrow}^{\dagger}\ d_{-k\downarrow})$$
and the matrix $\hat{H}_k$ is
\begin{widetext}
\begin{equation}
\hat{H}_k = \left( \begin{array}{cccc}
\varepsilon'_k-\mu-Um & \Delta_k & \varepsilon_k & 0\\
\Delta_k & -(\varepsilon'_k-\mu-Um) & 0 & -\varepsilon_k\\
\varepsilon_k & 0 & \varepsilon'_k-\mu+Um & \Delta_k\\
0 & -\varepsilon_k & \Delta_k & -(\varepsilon'_k-\mu+Um)
\end{array}
\right),
\end{equation}
\end{widetext}
with
\begin{eqnarray}
\varepsilon_k & = & -2t(\cos k_x +\cos k_y),\\
\varepsilon'_k & = & -4t'\cos k_x \cos k_y - 2t''(\cos 2k_x +\cos 2k_y),\\
\Delta_k & = & V\Delta(\cos k_x -\cos k_y).
\end{eqnarray}
Throughout the paper the summation $\sum_k$ with or without a prime on it means
that $k$ is within the magnetic Brillouin zone (BZ) $-\pi<k_x\pm k_y\le \pi$
or the original BZ, and the quantities with a hat represent matrices.
Under a unitary transformation $\hat{\Psi}_k=\hat{U}_k\hat{\tilde{\Psi}}_k$
with the new operator
$$\hat{\tilde{\Psi}}_k^{\dagger}=(\tilde{d}_{k\uparrow}^{\dagger}\ 
\tilde{e}_{-k\downarrow}\ \tilde{e}_{k\uparrow}^{\dagger}\ 
\tilde{d}_{-k\downarrow}),$$
the Hamiltonian (\ref{HM}) can be diagonalized into
\begin{equation}
H=\sum'_k [\hat{\tilde{\Psi}}_k^{\dagger} \hat{\tilde{H}}_k \hat{\tilde{\Psi}}_k
+ 2(\varepsilon'_k-\mu)]+N(Um^2+V\Delta^2),\label{HD}
\end{equation}
where $\hat{\tilde{H}}_k$ is a diagonal matrix
\begin{equation}
\hat{\tilde{H}}_k = \left(\begin{array}{llll}
E_k^- & & &\\
& -E_k^- & &\\
& & E_k^+ &\\
& & & -E_k^+
\end{array}\right)
\end{equation}
with the quasiparticle energy bands
\begin{eqnarray}
E_k^{\pm} & = & \sqrt{(\xi_k^{\pm}-\mu)^2+\Delta_k^2},\\
\xi_k^{\pm} & = & \varepsilon'_k\pm \sqrt{\varepsilon_k^2+U^2m^2}.
\end{eqnarray}
Here the spectra $E_k^{\pm}$ show formally the SC pairing in the two
AF bands $\xi_k^{\pm}$, which are originated from the beginning single band
in the presence of the AF order.
The unitary matrix $\hat{U}_k$ is
\begin{widetext}
\begin{equation}
\hat{U}_k=\left[\begin{array}{rr}
\cos\phi_k\left(\begin{array}{rr}
\cos\theta_k^- & \sin\theta_k^- \\
-\sin\theta_k^- & \cos\theta_k^- \end{array}\right) &
\sin\phi_k\left(\begin{array}{rr}
\cos\theta_k^+ & \sin\theta_k^+ \\
-\sin\theta_k^+ & \cos\theta_k^+ \end{array}\right)\\
-\sin\phi_k\left(\begin{array}{rr}
\cos\theta_k^- & \sin\theta_k^- \\
-\sin\theta_k^- & \cos\theta_k^- \end{array}\right) &
\cos\phi_k\left(\begin{array}{rr}
\cos\theta_k^+ & \sin\theta_k^+ \\
-\sin\theta_k^+ & \cos\theta_k^+ \end{array}\right)
\end{array}\right]
\label{U}
\end{equation}
\end{widetext}
with $\phi_k$ and $\theta_k^{\pm}$ satisfying
\begin{eqnarray}
\cos 2\phi_k & = & Um/\sqrt{\varepsilon_k^2+U^2m^2},\label{cosphi}\\
\sin 2\phi_k & = & \varepsilon_k/\sqrt{\varepsilon_k^2+U^2m^2},\label{sinphi}\\
\cos 2\theta_k^{\pm} & = & (\xi_k^{\pm}-\mu)/E_k^{\pm},\label{costhe}\\
\sin 2\theta_k^{\pm} & = & -\Delta_k/E_k^{\pm}.\label{sinthe}
\end{eqnarray}
The details of diagonalization are given in Appendix A. 

For late use, we define the $4\times 4$ matrix of the Matsubara
Green's functions
\begin{equation}
\hat{G}(k,\tau)=-\langle T_\tau \hat{\Psi}_k(\tau)\hat{\Psi}_k^{\dagger}(0)\rangle.
\end{equation}
Its Fourier transform is easily obtained
\begin{equation} 
\hat{G}(k,i\omega_n)=\hat{U}_k
(i\omega_n-\hat{\tilde{H}}_k)^{-1}\hat{U}_k^{\dagger}.
\label{GM}
\end{equation}

The free energy is given by $F=-T\ln ({\rm Tr} e^{-H/T})+N(1+x)\mu$
with $x$: electron doping concentration. This leads to
\begin{eqnarray}
F & = & -2T\sum'_{k,\nu=+,-} \ln [2\cosh (E_k^{\nu}/2T)]\nonumber\\
& & +N(\mu x + Um^2+V\Delta^2).
\end{eqnarray}
For given doping $x$ and temperature $T$, the order parameters $m$ and $\Delta$
are determined by minimizing the free energy, where
the chemical potential $\mu$ is fixed to ensure the correct electron number.

\section{Fermi surface evolution, quasiparticle property and electronic
density of states}
\begin{figure}[ht]
\begin{center}
\includegraphics[width=8.4cm,height=6.cm,clip]{FS.eps}\\
\vskip 3mm
\hskip -2mm
\includegraphics[width=8.4cm,height=2.5cm,clip]{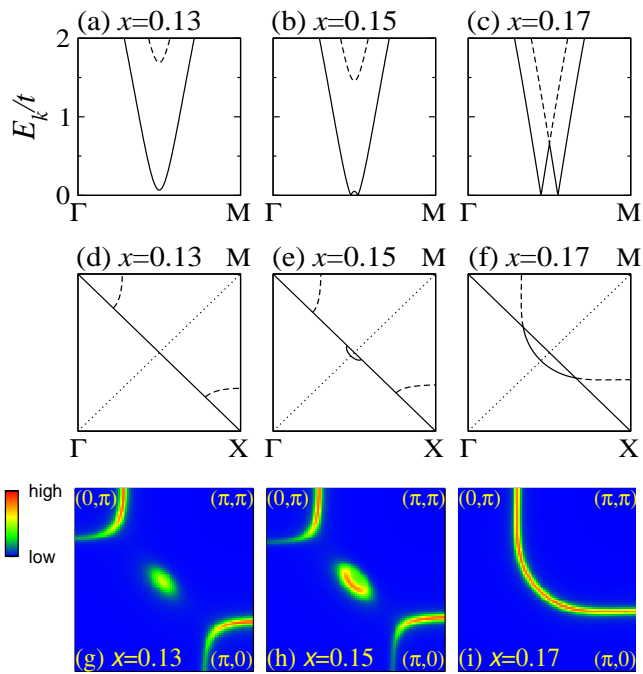}
\end{center}
\caption{(Color online) (a)-(c) Quasiparticle spectra $E_k^{\pm}$ at $T=0$ along the line $\Gamma(0,0)$--M$(\pi,\pi)$ and (d)-(f) Fermi surfaces for the corresponding
AF bands $\xi_k^{\pm}-\mu$ plotted in the first quadrant
[X=$(\pi,0)$]. In each panel the solid line is contributed by band ``$-$''
and the dashed one by ``$+$''. (g)-(i) Density plots of the corresponding
spectral intensity at $T=0$, obtained by integrating $A(k,\omega)$ [Eq.~(\ref{Eq:Akw})] over an energy interval $[-0.04,0]t$.
The broadening factor is $0.1t$.}
\label{Fig:FS}
\end{figure}

In this work we are focused on the SC (or SC+AF) phase
and choose three typical doping levels for study: $x=0.13,\ 0.15$ and $0.17$
which correspond to underdoping, optimal and overdoping, respectively.
In order to reproduce the experimental orders at these doping levels,
the model parameters $U$ and $V$ are slightly changed with $x$:
(i) $U=3.9,\ V=1$ at $x=0.13$ (ii) $U=3.8,\ V=1.1$ at $x=0.15$
(iii) $U=3.7,\ V=1.1$ at $x=0.17$.
The underlying strategy is to accept part experimental facts from the beginning
and based on them to work out other theoretical results for
experimental comparison. The calculated order parameters $m$ and 
$\Delta$ at $T=0$ are given in Fig.~\ref{Fig:PD}(b),
which show the experimental features including (i) the coexistence of both AF and SC orders at $x=0.13$ and $0.15$ (ii) the decreasing AF order with
increasing doping and the vanishing AF order at $x=0.17$ (iii) the strongest
SC order (i.e., biggest $\Delta$) at $x=0.15$.
Accordingly, the quasiparticle spectra $E_k^{\pm}$ at $T=0$
are plotted along the diagonal line $k_x=k_y$ (where $\Delta_k=0$)
in Figs.~\ref{Fig:FS}(a)-(c). It is realized that 
a minimal finite gap exists for $x=0.13$ but nodal excitations are
present around $(\pi/2,\pi/2)$ for $x\ge 0.15$.
To thoroughly understand these results, we have plotted the
Fermi surfaces for the corresponding AF bands $\xi_k^{\pm}-\mu$ in Figs.~\ref{Fig:FS}(d)-(f). Also, for direct comparison with the ARPES data,
we have calculated the electronic spectral function $A(k,\omega)$ 
(see below) and shown the intensity maps in Figs.~\ref{Fig:FS}(g)-(i). 
Explicitly, three different cases are distinguished:
\begin{enumerate}
\item At underdoping $x=0.13$ the FS is lacking around $(\pi/2,\pi/2)$ [Fig.~\ref{Fig:FS}(d)].
That is why the finite energy ($s$-wave-like) excitation appears despite
the $d$-wave SC pairing. At the same time, it should be emphasized that,
due to the proximity of the lower AF band to the Fermi level,
some spectral intensity is already contributed around $(\pi/2,\pi/2)$ [Fig.~\ref{Fig:FS}(g)], in consistence with the ARPES data \cite{Remark}
in Ref.~\onlinecite{Matsui94}.

\item At optimal doping $x=0.15$ the FS pocket around $(\pi/2,\pi/2)$ truly
forms [Fig.~\ref{Fig:FS}(e)]. Correspondingly, the spectral intensity becomes
enough strong around $(\pi/2,\pi/2)$ [Fig.~\ref{Fig:FS}(h)],
as revealed by ARPES in Ref.~\onlinecite{Armitage02}.
Also note that the FS pocket around $(\pi/2,\pi/2)$
is separated from the others around $(\pi,0)$ and $(0,\pi)$.

\item At overdoping $x=0.17$ a large FS emerges around $(\pi,\pi)$ [Fig.~\ref{Fig:FS}(f)]. With vanishing AF order, the two AF bands
will essentially merge into a single band and the several FS pieces
contributed by the two bands will recover to a large continuous curve.
At present the ARPES data on overdoped samples are scarce. According to
the evolutionary trend, the ARPES intensity map is expected
to look like Fig.~\ref{Fig:FS}(i).
\end{enumerate}
In both cases (ii) and (iii), gapless quasiparticles are induced
from the cross of the FS and the nodal line of the SC order parameter
(or SC gap) $\Delta_k$.

Above, the spectral function is calculated by $A(k,\omega)=(-1/\pi){\rm Im}
G(k,i\omega_n)|_{i\omega_n\rightarrow \omega+i0^+}$. $G(k,i\omega_n)$ is the
Fourier transform of the Matsubara Green's functions $G_{\sigma}(k,\tau)=
-\langle T_\tau c_{k\sigma}(\tau)c_{k\sigma}^{\dagger}(0)\rangle$,
which turns out to be spin $\sigma$-irrelevant. For details, one can see Appendix B.
$A(k,\omega)$ is derived as follows, for all $k$ within the original BZ,
\begin{eqnarray}
A(k,\omega) & = & {1\over 4}\sum_{\nu=+,-}
(1+\nu \sin 2\phi_k)[(1+\cos 2\theta_k^{\nu})\delta(\omega-E_k^{\nu}) \nonumber\\
& & \hspace*{2cm} +(1-\cos 2\theta_k^{\nu})\delta(\omega+E_k^{\nu})]. \label{Eq:Akw}
\end{eqnarray}

\begin{figure}[ht]
\begin{center}
\includegraphics[width=8.4cm,height=4cm,clip]{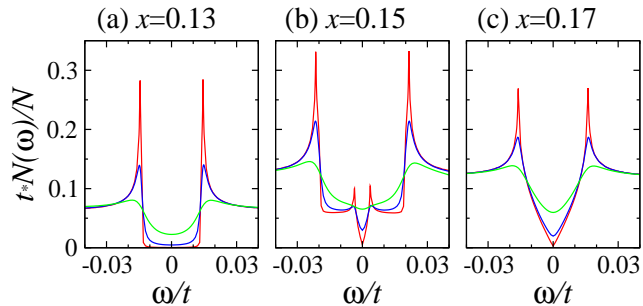}
\end{center}
\caption{(Color online) Electronic density of states at low energy
(in unit of $t\sim 300$ meV) and $T=0$. In each panel, the red (dark gray),
blue (black) and green (gray) curves show the results under the broadening factors $\Gamma/t=10^{-4}$, $10^{-3}$ and $5\times 10^{-3}$, respectively.}
\label{Fig:Nw}
\end{figure}

Furthermore, the electronic DOS per spin is given by
$N(\omega)=\sum_k A(k,\omega)$,
which is simplified into
\begin{eqnarray}
N(\omega) & = & {1\over 4} \sum_{k,\nu=+,-} 
[(1+\cos 2\theta_k^{\nu})\delta(\omega-E_k^{\nu}) \nonumber\\
& & \hspace*{1.5cm} +(1-\cos 2\theta_k^{\nu})\delta(\omega+E_k^{\nu})].
\end{eqnarray}
The electronic DOS at low energy is plotted in Fig.~\ref{Fig:Nw},
where the results obtained with different broadening factors $\Gamma$ are given.
Let us focus on the curves with $\Gamma/t=10^{-4}$ [red (dark gray) lines].
As expected, the DOS takes on a ``U'' shape at $x=0.13$ indicating a
finite energy gap, whereas it has a ``V'' shape around zero energy
for $x\ge 0.15$ showing vanishing gap.

At this stage, we would like to point out another remarkable finding in the DOS,
i.e., the double-peak structure in the case of $x=0.15$.
(Only the half side $\omega>0$ is talked about.)
Careful investigation shows that the small peak at lower energy is the coherence peak for the ``$-$'' band, i.e., the peak position corresponds to the SC gap maximum opened on the small FS piece around $(\pi/2,\pi/2)$
[solid line in Fig.~\ref{Fig:FS}(e)], and the large one at higher energy is the coherence peak for the ``$+$'' band, i.e., the peak position corresponds to that
on the FS pieces shown by the dashed lines in Fig.~\ref{Fig:FS}(e).
Thus this double-peak structure is a characteristic feature for the
e-doped cuprates at optimal doping which have a segmented FS and for the two-band modeling of them. In contrast, at $x=0.17$ the two bands recover to the single band and the FS turns into a large continuous curve. Then only a
single coherence peak shows up [Fig.~\ref{Fig:Nw}(c)], at the energy
corresponding to the SC gap maximum on this whole large FS.
So far, the double-peak structure has not yet been observed in tunneling experiments on optimally doped NCCO.\cite{Kashiwaya,Shan}
One possible reason is the effect of the broadening factor $\Gamma$.
As seen from Fig.~\ref{Fig:Nw},
with increasing $\Gamma$ the sharp peaks will be smoothened and in particular,
the small peak at the lower energy in the case of $x=0.15$ will
be smeared out. Actually, for $\Gamma=5\times 10^{-3}t\sim 1.5$ meV which is in the order of the corresponding experimental value,\cite{Shan} the DOS shape
at $x=0.15$ is more suggestive of a $s$-wave-like gap.
Apparently, much smaller $\Gamma$ is required in order to observe the double-peak structure and further experiments are needed to clarify the issue.

After extracting the properties of the quasiparticle excitation and
electronic DOS at different doping levels, we continue to study
their consequences to the physical quantities in the following.

\section{Heat capacity}
First we calculate the heat capacity, which is given by
$C=T(\partial S/\partial T)$ with the entropy $S=-\partial F/\partial T$.
Explicitly $S$ is expressed as
\begin{equation}
S=2\sum_{k,\nu=+,-}' \left\{\ln [2\cosh (E_k^{\nu}/2T)]-
(E_k^{\nu}/2T)\tanh (E_k^{\nu}/2T)\right\}.
\end{equation}
For general $T$, the derivative $\partial S/\partial T$ has no explicit
form due to the $T$-dependence of the order parameters.
Only at low $T\ll T_c$ (the SC transition temperature), the order parameters
are nearly $T$-independent, leading to the simple formula for $C$, i.e.,
\begin{equation} 
C=(1/2T^2)\sum_{k,\nu=+,-}' (E_k^{\nu})^2/\cosh^2 (E_k^{\nu}/2T).
\end{equation}

\begin{figure}[ht]
\begin{center}
\includegraphics[width=8.4cm,height=6.cm,clip]{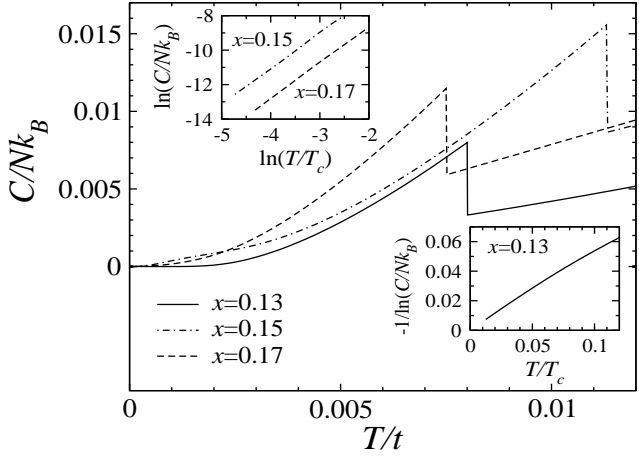}
\end{center}
\caption{Specific heat as a function of temperature at three doping levels.
It is seen $T_c/t=0.008,\ 0.0113$ and $0.0075$ at $x=0.13,\ 0.15$ and $0.17$,
respectively. The data at low temperatures ($T/T_c\ll 1$) are replotted in the lower inset for $x=0.13$ and upper inset for $x=0.15$ and $0.17$.
Note that the different axis rescales are used in the two insets.}
\label{Fig:cap}
\end{figure}

The heat capacity per site is shown in Fig.~\ref{Fig:cap}.
The obtained transition temperatures at three doping levels are around $30$ K,
close to the experimental data. That the $C$ values at $x=0.15$ are bigger
than those at $x=0.17$ in the region about $T/t<0.002$ is consistent
with the fact seen in Fig.~\ref{Fig:Nw} that the DOS at $x=0.15$
is greater than that at $x=0.17$ in the very low
energy region [below the small peak in Fig.~\ref{Fig:Nw}(b)].
Here we are mainly interested in the qualitative behaviors
at low $T\ll T_c$. It is found that the heat capacity takes on an exponential
$T$-dependence for $x=0.13$ due to the $s$-wave-like excitations,
but a power law behavior for $x=0.15$ and $0.17$ due to the nodal excitations.
To more clearly see this point, we have replotted the low-$T$ data for $x=0.13$ in the lower inset by using $-1/\ln (C/Nk_B)$ vs $T/T_c$ and those for $x=0.15$ and $0.17$ in the upper inset with logarithmic scales for both axes. Now we have nearly straight lines in all cases. Moreover, it is easy to check that the slopes for the two lines in the upper inset are both $2$, which means $T^2$ dependence of $C$ for $x=0.15$ and $0.17$.
In addition, we find that in the pure AF phase without SC order, e.g., at doping $x=0.1$, the heat capacity takes on a linear $T$-dependence (not shown), indicating an AF metal. So it is of great interest to experimentally extract the electronic part of the heat capacity and investigate the variation of its 
$T$-dependence with doping.

\section{Superfluid density}
Another important quantity for calculation is the superfluid density $\rho$
which relates to the penetration depth $\lambda$ by $\rho\propto \lambda^{-2}$.
The superfluid density divided by the carrier effective mass $m^*$ is given by \cite{Scalapino}
\begin{equation}
\rho/m^*= \langle -K_x \rangle /N - 
\Lambda (q_x=0,q_y\rightarrow 0,i\Omega_n=0),\label{rho}
\end{equation}
where the operator $K_x$ is derived as follows
($\vec{x},\ \vec{y}$: unit vectors in $x,\ y$ directions, respectively)
\begin{eqnarray}
K_x & = & -t\sum_{l\sigma} (c_{l+\vec{x}\sigma}^{\dagger}c_{l\sigma}+
{\rm H.c.})-t'\sum_{l\sigma,\delta'=\vec{x}\pm \vec{y}} (c_{l+\delta'\sigma}^{\dagger}c_{l\sigma}+{\rm H.c.})\nonumber\\
& & -4t''\sum_{l\sigma} (c_{l+2\vec{x}\sigma}^{\dagger}c_{l\sigma}+{\rm H.c.})
\end{eqnarray}
and the current-current correlation function $\Lambda$ is ($\beta=1/T$ and
$\Omega_n=2n\pi/\beta$: bosonic Matsubara frequencies)
\begin{equation}
\Lambda(q,i\Omega_n)={1\over N}\int_0^{\beta} {\rm d}\tau e^{i\Omega_n\tau}
\langle T_{\tau}J_q^x(\tau)J_{-q}^x(0)\rangle\label{Lambda}
\end{equation}
with $J_q^x=\sum_l e^{-i\vec{q}\cdot\vec{R}_l}J_l^x$. Here
$J_l^x$ is the $x$-component of the paramagnetic current operator
at site $l$
\begin{eqnarray}
J_l^x & = & it\sum_{\sigma} (c_{l+\vec{x}\sigma}^{\dagger}c_{l\sigma}-
{\rm H.c.})+it'\sum_{\sigma,\delta'=\vec{x}\pm \vec{y}} (c_{l+\delta'\sigma}^{\dagger}c_{l\sigma}-{\rm H.c.})\nonumber\\
& & +2it''\sum_{\sigma} (c_{l+2\vec{x}\sigma}^{\dagger}c_{l\sigma}-{\rm H.c.}).
\end{eqnarray}

The first term of Eq.~(\ref{rho}) can be easily calculated if
$K_x$ is expanded in momentum space, i.e.,
\begin{equation}
K_x =\sum'_k \hat{\Psi}_k^{\dagger} \hat{P}_k \hat{\Psi}_k=
\sum'_k \hat{\tilde{\Psi}}_k^{\dagger} \hat{\tilde{P}}_k \hat{\tilde{\Psi}}_k,
\end{equation}
with
\begin{equation}
\hat{P}_k = \left(\begin{array}{cccc}
g'_k & 0 & g_k & 0\\
0 & -g'_k & 0 & -g_k\\
g_k & 0 & g'_k & 0\\
0 & -g_k & 0 & -g'_k
\end{array}\right),\ 
\hat{\tilde{P}}_k=\hat{U}_k^{\dagger}\hat{P}_k\hat{U}_k,
\end{equation}
$g_k=-\partial^2\varepsilon_k/ \partial k_x^2$ and
$g'_k = -{\partial^2\varepsilon'_k/ \partial k_x^2}$.
Then the thermal average is obtained
\begin{equation}
\langle K_x\rangle=\sum'_k  \sum_{s=1}^4 (\hat{\tilde{P}}_k)_{ss}/(1+e^{E_k^s/T}),
\end{equation}
where $(\hat{\tilde{P}}_k)_{ss'}$ represents the matrix element of $\hat{\tilde{P}}_k$ and for convenient statement the quasiparticle 
energy bands are relabeled by $E_k^s\ (s=1,2,3,4)$
with $E_k^1=E_k^-,\ E_k^2=-E_k^-,\ E_k^3=E_k^+$ and $E_k^4=-E_k^+$.

To calculate the second term of Eq.~(\ref{rho}), we first write down
the current operator $J_q^x$, i.e.,
\begin{equation}
J_q^x=\sum'_k \hat{\Psi}_{k-q}^{\dagger}\hat{V}_{k,q}\hat{\Psi}_k,
\end{equation}
with
\begin{equation}
\hat{V}_{k,q}= \left(\begin{array}{cccc}
f'_{k,q} & 0 & f_{k,q} & 0\\
0 & -f'_{-k+q,q} & 0 & -f_{-k+q,q}\\
f_{k,q} & 0 & f'_{k,q} & 0\\
0 & -f_{-k+q,q} & 0 & -f'_{-k+q,q}
\end{array}\right).
\end{equation}
Above, $f_{k,q} = it[e^{-i(k_x-q_x)}-e^{ik_x}]$ and
$f'_{k,q} = 2it'[e^{-i(k_x-q_x)}\cos (k_y-q_y)-e^{ik_x}\cos k_y]
+2it''[e^{-2i(k_x-q_x)}-e^{2ik_x}]$,
which reduce to the simple form at $q=0$: 
$f_{k,0}=\partial \varepsilon_k/ \partial k_x$
and $f'_{k,0}=\partial \varepsilon'_k/ \partial k_x$.
According to Eq.~(\ref{Lambda}), the correlation function $\Lambda$ is
derived as, after expansion of $\hat{G}(k,\tau)$ into $\hat{G}(k,i\omega_n)$,
\begin{eqnarray}
\Lambda(q,i\Omega_n) & = & -{1\over N\beta}\sum'_k\sum_{i\omega_n}
{\rm Tr}[\hat{G}(k-q,i\omega_n)\hat{V}_{k,q} \nonumber\\
& & \hspace*{1cm} \hat{G}(k,i\Omega_n+i\omega_n)\hat{V}_{k-q,-q}]. 
\end{eqnarray}
Substituting $\hat{G}(k,i\omega_n)$ with Eq.~(\ref{GM}) and completing
the summation over $i\omega_n$, we obtain
\begin{eqnarray}
\Lambda(q,i\Omega_n) & = & {1\over 2N}\sum'_k \sum_{s,s'=1}^4 (\hat{W}_{k,q})_{ss'}(\hat{W}_{k-q,-q})_{s's}\nonumber\\
& & \times {\tanh (E_{k-q}^s/2T)-\tanh (E_k^{s'}/2T) \over 
i\Omega_n+E_{k-q}^s-E_k^{s'}},
\end{eqnarray}
where $(\hat{W}_{k,q})_{ss'}$ represents the matrix element of $\hat{W}_{k,q}=\hat{U}_{k-q}^{\dagger}\hat{V}_{k,q}\hat{U}_k$.
Finally we set $i\Omega_n=0$ and then take the limit $q_y\rightarrow 0$
to have
\begin{widetext}
\begin{equation}
\Lambda(q_x=0,q_y\rightarrow 0,i\Omega_n=0) =  
{1\over 2N}\sum'_k\sum_{s=1}^4 \left\{
\sum_{s'\neq s} [(\hat{W}_{k,0})_{ss'}]^2
{\tanh (E_k^s/2T)-\tanh (E_k^{s'}/2T) \over E_k^s-E_k^{s'}}+
{[(\hat{W}_{k,0})_{ss}]^2 \over 2T\cosh^2(E_k^s/2T)}
\right\}.
\end{equation}
\end{widetext}

\begin{figure}[ht]
\begin{center}
\includegraphics[width=8.4cm,height=6cm,clip]{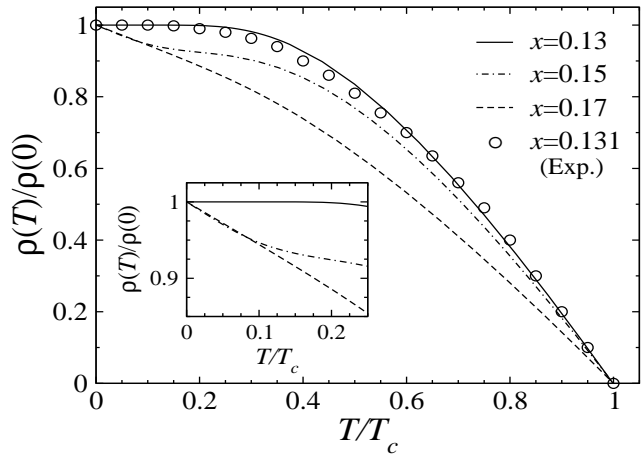}
\end{center}
\caption{Superfluid density as a function of temperature at three doping 
levels. The low-$T$ regime is enlarged in the inset.
The circles in the main panel are the experimental data for PCCO at $x=0.131$
extracted from Ref.~\onlinecite{Kim} (see text).}
\label{Fig:fluid}
\end{figure}

The superfluid density is given in Fig.~\ref{Fig:fluid},
where the ratio $\rho(T)/\rho(0)$ vs $T/T_c$ is plotted.
Still we are concentrated on the low-$T$ regime, which is enlarged in the inset.
The notable observation is that the curve for $x=0.13$ is very flat, which may be characterized by the formula $1-\rho(T)/\rho(0)\propto \exp (-b/T)$ with
a constant $b$. When the quasiparticle excitations become gapless at
$x=0.15$ and $0.17$, the results are much different. Now $\rho(T)$ decreases
faster, with a linear relation $1-\rho(T)/\rho(0)\propto T$,
as apparently seen from $x=0.17$ in the inset of Fig.~\ref{Fig:fluid}.
At $x=0.15$, $\rho(T)/\rho(0)$ in the main panel shows an inflexion point as a result of crossover from $x=0.13$ to $x=0.17$. Similar to the case for $x=0.17$,
the linear relation holds at $T/T_c<0.1$ due to the nodal excitations on the
small FS pocket around $(\pi/2,\pi/2)$ [solid line in Fig.~\ref{Fig:FS}(e)].
When $T$ increases the low energy quasiparticles
around this small FS pocket are completely excited. Then the change
of the superfluid density is controlled by the finite energy excitations
around the FS pockets circling $(\pi,0)$ and $(0,\pi)$, similar to
the case for $x=0.13$. So the curve at $x=0.15$ away from low-$T$
tends to behave like the one at $x=0.13$.
The linear low-$T$-dependence at $x=0.15$ and $0.17$ may become $T^2$
in real materials if impurity scattering is considered.\cite{Hirschfeld}

The low-$T$-dependence of the superfluid density (or penetration depth)
is important to hint the SC pairing symmetry and
has been substantially measured for e-doped cuprates.\cite{Alff,Kim,Kokales,Snezhko,Skinta}
However, the experimental results are controversial and some of them might
be misexplained. For example, the $s$-wave SC gap was inferred, from the flat feature of the experimental data, by Kim {\it et al.} \cite{Kim} for PCCO over a wide doping range $0.124\le x\le 0.144$. In Fig.~\ref{Fig:fluid} our theoretical result at $x=0.13$ is compared with their data at $x=0.131$, and good agreement
is obtained. But actually in our theory the SC gap $\Delta_k$ is $d$-wave. It should be reminded that the doping level $x=0.131$ in the work by Kim {\it et al}. corresponds to the optimal doping, although this optimal value is $0.15$ for PCCO in most experiments.\cite{Dagan,Zimmers,Tsuei,Snezhko} Our result suggests that,
even with a $d$-wave SC pairing, it is not difficult to understand the $s$-wave-like excitation from some penetration depth measurements at various doping levels \cite{Alff,Kim,Skinta} if nominally optimally or overdoped samples in those measurements are in reality underdoped, or more generally,
if the FS is absent around $(\pi/2,\pi/2)$ in those samples perhaps due to
the still strong AF order. On the other hand, the power law
$T$-dependence ($T$ or $T^2$) of $\rho$
at optimal and overdoping is consistent with other penetration depth
measurements \cite{Kokales,Snezhko} and many others such as phase sensitive Josephson junctions \cite{Tsuei,Ariando} and ARPES.\cite{Armitage01}
Thus in this way the diverse experimental results could be largely reconciled.
Finally, we mention that our essential idea to exhibit the $s$-wave-like excitation and to explain the flat feature of some penetration depth data is different from the phenomenological superfluid response theory in
Ref.~\onlinecite{Luo} where the existence of nodal quasiparticle excitations
is assumed at all doping levels.

\section{Conclusion}
In conclusion, we have explored an intrinsic physical mechanism,
based on the FS evolution with doping, to understand
the diverse experimental results on the SC pairing symmetry in e-doped cuprates.
We argue that the FS pocket around $(\pi/2,\pi/2)$ has not yet formed until doping reaches about the optimal value. Therefore, in the underdoped regime, even if the SC pairing order parameter is $d$-wave which vanishes along the diagonal line, the quasiparticle excitation is still gapped which shows a $s$-wave-like behavior due to the absence of the FS across that line.
To demonstrate this idea, we have elaborated a theory with consideration of
coexisting AF and SC orders in e-doped cuprates.
The variations of the quasiparticle property and electronic DOS are exhibited
as the FS evolves from underdoping to overdoping.
The physical quantities heat capacity and superfluid density are calculated and their low temperature dependences are compared at different doping levels. Our study gives new understanding of some measurements which actually detected the $s$-wave character of the quasiparticle excitation (rather than the SC order parameter) and thus tends to reconcile the apparently contradictory experimental
results. We have also predicted the remarkable double-peak feature in the DOS
for the e-doped cuprates at optimal doping which have a segmented FS.

\section*{ACKNOWLEDGMENTS}
This work was supported by the Texas Center for Superconductivity at
the University of Houston and the Robert A. Welch Foundation under
Grant No. E-1146.

\appendix
\section{Hamiltonian diagonalization}
We consider the following two steps to diagonalize the Hamiltonian (\ref{HM}),
which is explicitly
\begin{eqnarray}
H & = & \sum'_{k\sigma} \varepsilon_k (d_{k\sigma}^{\dagger}e_{k\sigma}+{\rm H.c.})
+\sum'_{k\sigma}(\varepsilon'_k-\mu-\sigma Um)d_{k\sigma}^{\dagger}d_{k\sigma}
\nonumber\\
& & +\sum'_{k\sigma}(\varepsilon'_k-\mu+\sigma Um)e_{k\sigma}^{\dagger}e_{k\sigma}
\nonumber\\
& & -\sum'_k \Delta_k (d_{k\uparrow}e_{-k\downarrow}+e_{k\uparrow}d_{-k\downarrow}+
{\rm H.c.})\nonumber\\
& & +N(Um^2+V\Delta^2). \label{Hk}
\end{eqnarray}

First, we use the rotational transformations to remove the 
hopping term between $d$ and $e$, i.e.,
\begin{equation}
\left( \begin{array}{l}
d_{k\uparrow}\\
e_{k\uparrow}\end{array} \right)= \hat{U}_{1k}
\left( \begin{array}{l}
\bar{d}_{k\uparrow}\\
\bar{e}_{k\uparrow}\end{array} \right),\ 
\left( \begin{array}{l}
e_{k\downarrow}\\
d_{k\downarrow}\end{array} \right)= \hat{U}_{1k}
\left( \begin{array}{l}
\bar{e}_{k\downarrow}\\
\bar{d}_{k\downarrow}\end{array} \right)\label{U1}
\end{equation}
with [see Eqs.~(\ref{cosphi}) and (\ref{sinphi}) for the choice of $\phi_k$]
\begin{equation}
\hat{U}_{1k}=\left( \begin{array}{rr}
\cos \phi_k & \sin \phi_k\\
-\sin \phi_k & \cos \phi_k \end{array} \right).
\end{equation}
In terms of the new operators $\bar{d}$ and $\bar{e}$, (\ref{Hk}) turns into
\begin{eqnarray}
H & = & \sum'_k (\xi_k^- -\mu) (\bar{d}_{k\uparrow}^{\dagger}\bar{d}_{k\uparrow}
+\bar{e}_{k\downarrow}^{\dagger}\bar{e}_{k\downarrow})\nonumber\\
& & +\sum'_k (\xi_k^+ -\mu) (\bar{e}_{k\uparrow}^{\dagger}\bar{e}_{k\uparrow}
+\bar{d}_{k\downarrow}^{\dagger}\bar{d}_{k\downarrow}) \nonumber\\
& & -\sum'_k \Delta_k (\bar{d}_{k\uparrow}\bar{e}_{-k\downarrow}+
\bar{e}_{k\uparrow}\bar{d}_{-k\downarrow}+{\rm H.c.})\nonumber\\
& & +N(Um^2+V\Delta^2),\label{Hkb}
\end{eqnarray}
where the pairing term is formally unchanged.

Second, we adopt the Bogliubov transformations
\begin{equation}
\left( \begin{array}{l}
\bar{d}_{k\uparrow}\\
\bar{e}_{-k\downarrow}^{\dagger}\end{array} \right)= \hat{U}_{2k}^-
\left( \begin{array}{l}
\tilde{d}_{k\uparrow}\\
\tilde{e}_{-k\downarrow}^{\dagger}\end{array} \right),\ 
\left( \begin{array}{l}
\bar{e}_{k\uparrow}\\
\bar{d}_{-k\downarrow}^{\dagger}\end{array} \right)= \hat{U}_{2k}^+
\left( \begin{array}{l}
\tilde{e}_{k\uparrow}\\
\tilde{d}_{-k\downarrow}^{\dagger}\end{array} \right)\label{U2}
\end{equation}
with [see Eqs.~(\ref{costhe}) and (\ref{sinthe}) for the choice of $\theta_k^{\pm}$]
\begin{equation}
\hat{U}_{2k}^{\pm} =
\left( \begin{array}{rr}
\cos \theta_k^{\pm} & \sin \theta_k^{\pm}\\
-\sin \theta_k^{\pm} & \cos \theta_k^{\pm} \end{array} \right).
\end{equation}
Then (\ref{Hkb}) is diagonalized into
\begin{eqnarray}
H & = & \sum'_k [E_k^- (\tilde{d}_{k\uparrow}^{\dagger}\tilde{d}_{k\uparrow}-
\tilde{e}_{-k\downarrow}\tilde{e}_{-k\downarrow}^{\dagger})\nonumber\\
& & +E_k^+ (\tilde{e}_{k\uparrow}^{\dagger}\tilde{e}_{k\uparrow}-
\tilde{d}_{-k\downarrow}\tilde{d}_{-k\downarrow}^{\dagger})
+2(\varepsilon'_k-\mu)]\nonumber\\
& & +N(Um^2+V\Delta^2),
\end{eqnarray}
which is exactly the Hamiltonian (\ref{HD}). If all the transformations
(\ref{U1}) and (\ref{U2}) are integrated, we will arrive at the global transformation $\hat{\Psi}_k=\hat{U}_k\hat{\tilde{\Psi}}_k$
with $\hat{U}_k$ given by Eq.~(\ref{U}).

\section{Derivation of the spectral function Eq.~(\ref{Eq:Akw})}
With division of the original lattice into two sublattices,
one has the following correspondence for the operators $c$ and $d/e$
in momentum space [$Q=(\pi,\pi)$]
\begin{eqnarray}
c_{k\sigma} & = & (d_{k\sigma}+e_{k\sigma})/\sqrt{2},\\
c_{k+Q\sigma} & = & (d_{k\sigma}-e_{k\sigma})/\sqrt{2},
\end{eqnarray}
where $k$ is restricted into the magnetic BZ.
Then the Green's functions for the original $c$-electron
can be expressed as follows, with help of the matrix elements [denoted by $\hat{G}_{ss'}(k,\tau)$] of $\hat{G}(k,\tau)$,
\begin{eqnarray}
G_{\uparrow}(k,\tau) & = & {1\over 2} \sum_{s,s'=1,3} \hat{G}_{ss'}(k,\tau),\\
G_{\downarrow}(k,\tau) & = & -{1\over 2} \sum_{s,s'=2,4} \hat{G}_{ss'}(-k,-\tau),\\
G_{\uparrow}(k+Q,\tau) & = & {1\over 2} \sum_{s,s'=1,3} \epsilon_{ss'} \hat{G}_{ss'}(k,\tau),\\
G_{\downarrow}(k+Q,\tau) & = & -{1\over 2} \sum_{s,s'=2,4} \epsilon_{ss'}
\hat{G}_{ss'}(-k,-\tau),
\end{eqnarray}
where the coefficient $\epsilon_{ss'}=1\ (-1)$ for $s'=s$\ ($s'\neq s$)
is introduced. All the above equations are also valid for the Fourier transformed functions after the replacement $\tau\rightarrow i\omega_n$.

By use of Eq.~(\ref{GM}), we can obtain the following sums
\begin{widetext}
\begin{eqnarray}
I_1(k,i\omega_n) & = & \sum_{s,s'=1,3}\hat{G}_{ss'}(k,i\omega_n)
 \hspace*{5mm} =  \sum_{\nu=+,-}
(1+\nu\sin 2\phi_k)\left({\cos^2\theta_k^{\nu} \over i\omega_n-E_k^{\nu}}+
{\sin^2\theta_k^{\nu} \over i\omega_n+E_k^{\nu}}\right),\\
I_2(k,i\omega_n) & = & \sum_{s,s'=2,4}\hat{G}_{ss'}(k,i\omega_n)
 \hspace*{5mm} =  \sum_{\nu=+,-}
(1+\nu\sin 2\phi_k)\left({\sin^2\theta_k^{\nu} \over i\omega_n-E_k^{\nu}}+
{\cos^2\theta_k^{\nu} \over i\omega_n+E_k^{\nu}}\right),\\
I_3(k,i\omega_n) & =  & \sum_{s,s'=1,3}\epsilon_{ss'} \hat{G}_{ss'}(k,i\omega_n)
  =  \sum_{\nu=+,-}
(1-\nu\sin 2\phi_k)\left({\cos^2\theta_k^{\nu} \over i\omega_n-E_k^{\nu}}+
{\sin^2\theta_k^{\nu} \over i\omega_n+E_k^{\nu}}\right),\\
I_4(k,i\omega_n) & = & \sum_{s,s'=2,4}\epsilon_{ss'}\hat{G}_{ss'}(k,i\omega_n)
  =  \sum_{\nu=+,-}
(1-\nu\sin 2\phi_k)\left({\sin^2\theta_k^{\nu} \over i\omega_n-E_k^{\nu}}+
{\cos^2\theta_k^{\nu} \over i\omega_n+E_k^{\nu}}\right).
\end{eqnarray}
It is easy to check $G_{\downarrow}(k,i\omega_n)=-{1\over 2} I_2(-k,-i\omega_n)=-{1\over 2}I_2(k,-i\omega_n)={1\over 2} I_1(k,i\omega_n)
=G_{\uparrow}(k,i\omega_n)$. Similarly
one has $G_{\downarrow}(k+Q,i\omega_n)=G_{\uparrow}(k+Q,i\omega_n)$.
So the spin index can be dropped out and we obtain
\begin{eqnarray}
G(k,i\omega_n) & = & {1\over 2}I_1(k,i\omega_n),\label{GS1}\\
G(k+Q,i\omega_n) & = & {1\over 2} I_3(k,i\omega_n)=
{1\over 2} I_1(k+Q,i\omega_n),
\end{eqnarray}
\end{widetext}
where the last equation holds if the definitions for $\phi_k$, $\theta_k^{\pm}$
and $E_k^{\pm}$ are extended to the original BZ.
Therefore, no matter $k$ is within or beyond the magnetic BZ,
$G(k,i\omega_n)$ has the unified form (\ref{GS1}).
Analytical continuation $i\omega_n\rightarrow \omega+i0^+$ for $G(k,i\omega_n)$
will lead to Eq.~(\ref{Eq:Akw}).


\begin{references}
\bibitem{Remark} The ARPES intensity map for NCCO at $x=0.13$ in Ref.~\onlinecite{Matsui94} was obtained at $T=30$ K, i.e., in the normal state
(but with AF order). We believe that the similar map is valid at lower $T$ 
where the compound is in the SC+AF state as we consider here.
\end{references}
\end{document}